\documentclass {JHEP3}

\newcommand{\be}{\begin{equation}}
\newcommand{\ee}{\end{equation}}
\newcommand{\bea}{\begin{eqnarray}}
\newcommand{\eea}{\end{eqnarray}}

\def\fr{\frac}

\def\l{\lambda}

\def\s{\sigma}
\def\S{\Sigma}
\def\t{\tau}

\def\O{\Omega}

\newcommand{\ba}{\begin{array}}
\newcommand{\ea}{\end{array}}

\let\la=\label
\let\bm=\bibitem

\author{ Nihat Sadik Deger$^{1,3}$ and Ali Kaya$^{2,3}$\\
$^1$ Department of Mathematics, Bogazici University, Bebek, 34342, Istanbul-Turkey \\
$^2$ Department of Physics, Bogazici University, Bebek, 34342, Istanbul-Turkey \\
$^3$ Feza Gursey Institute, Cengelkoy, 34680, Istanbul-Turkey \\

E-mails: \email{sadik.deger@boun.edu.tr}, \email{ali.kaya@boun.edu.tr}
}

\abstract{We construct an intersecting S-brane solution of 11-dimensional supergravity for which the contribution of the Chern-Simons term to the field equations is non-zero. 
After studying some of its properties, we consider three different compactifications (each with 3 separate subcases) of this system to 4-dimensions. Two of these give accelerating cosmologies, however their expansion factors are of order unity. We also find two static versions of this configuration and its dimensional reduction to type IIA theory.}

\title{Chern-Simons S-Brane Solutions in M-theory and Accelerating Cosmologies}

\begin{document}


\section{Introduction}

In the $D=11$ supergravity \cite{11} the field equation of the 4-form field strength $F$ looks like
$$d*F \sim F\wedge F$$
where the right-hand-side comes from the Chern-Simons term in the action.
For basic $p$-brane solutions of this theory, namely the membrane \cite{2brane} and the five brane \cite{5brane}, $F\wedge F$ vanishes automatically. This happens also when intersections \cite{inter1, inter2, inter3} are considered. Actually, investigating structure of the 4-forms of an $M2$ and an 
$M5$-brane it is clear that it is impossible to make $F\wedge F$ nonzero and satisfy the above equation by additional branes. This is because 4-forms of $M2$'s have two common directions (the time coordinate and the radial coordinate of the overall transverse space), whereas $M5$-brane 4-form has none of these components. However, relaxing the condition to have only basic $M$-branes one can have {\sl composite M-brane}
solution \cite{dyon1, dyon2, dyon3}. This is obtained using the U-duality transformations and it is half supersymmetric. Another way to construct such solutions for an $M2$-brane is to replace its 8-dimensional flat transverse part with a Ricci-flat space that supports a non-trivial harmonic 4-form. This modifies the 4-form ansatz of the membrane and the Chern-Simons term becomes active \cite{harmonic1} - \cite{harmonic7}, which sometimes resolves the singularity of the $M2$-brane solution.

The situation is similar for spacelike 
branes; the Chern-Simons term plays no role in
SM2 and SM5 branes \cite{pope1,s1, s2, s3}. Meanwhile, in intersections 
branes are located so that $F\wedge F$ is either trivially zero \cite{sinter1} or charges are chosen proportional so that it vanishes when all terms are added up \cite{nonstandard}. However, unlike $p$-brane intersections it is possible to have a non-zero Chern-Simons term and satisfy the 4-form field equation simultaneously. This is mainly due to the fact that one can write down 4-forms of an SM2 and an SM5-brane without any overlapping and SM2-brane 4-forms must have only the time coordinate in common. The minimal configuration contains two SM2 and one SM5-brane and it is unique up to renaming of coordinates. This case was considered in \cite{nonstandard}, but field equations could be analyzed only numerically.
In this paper we find an analytic solution for this system and investigate its properties in section 2.
To the best of our knowledge this is the first example of such a time-dependent solution. 

SM2-branes upon compactification can give rise to 4-dimensional accelerating cosmologies \cite{acc1}-\cite{acc5}
(for earlier work on S-brane cosmology see e.g. \cite{old1,old2}). The second goal of this paper is to see whether this is still true and if so whether there is any improvement,
when the Chern-Simons term is active. In section 3 we consider three different compactifications (each with 3 distinct subcases) to $D=4$ and find that in this respect there is not much difference;
there is acceleration in two of these, however like usual SM2-branes the number of e-foldings is order 1. 

After these, we find two static versions of our solution in section 4 and discuss some of their properties. We conclude in section 5 with some comments and future directions.

\section{The Solution}

Here we present a detailed  construction of the Chern-Simons S-brane solution and discuss its basic properties. As we will see, compared to previously obtained S-brane solutions there are crucial differences in field equations, and methods that have been employed for solving them do not work anymore. Specifically, in our case the 4-form field equation is not satisfied identically and there appear two first order equations to be worked out. Moreover, it is not possible to decouple the differential equations like usual S-branes. Instead, we find a suitable ansatz that helps us to solve the system step by step in a consistent manner. Readers who are more interested with the solution itself may skip the derivation and go directly to the subsection \ref{basic}.

\subsection{Derivation}

The bosonic action of the 11-dimensional supergravity \cite{11} can be written as
\be
S=\int d^{11}x(\sqrt{-g}R-\frac{1}{2} F \wedge *F + \frac{1}{6}
F\wedge F \wedge A) \, ,
\ee
where the last term is the Chern-Simons term. The equations of motion are given by
\bea\label{eins1}
R_{AB}&=&\frac{1}{2.3!}F_{ACDE}F_{B}{}^{CDE}-\frac{1}{6.4!}g_{AB}F^{2},\\
d*F&=&\frac{1}{2}F\wedge F \, .
\label{eins2}
\eea
We also have the Bianchi identity $dF=0$. 

Looking at 4-form field strengths of SM2 and SM5 branes together with (\ref{eins2}) we see that at least 3 branes are needed to achieve a non-trivial $F \wedge F$ contribution. This is unique up to relabeling of coordinates. The configuration that we will discuss below was first considered in \cite{nonstandard} where an analytic solution couldn't be obtained. However, using numerical techniques the behavior of the metric functions were studied and it was observed that their asymptotic values do not depend much on initial conditions and they always approach to zero as $t\rightarrow \infty$ which signals a singularity.\footnote{We verify that graphs given in  \cite{nonstandard} correspond to the solution that is presented in this section with a specific choice of integration constants.}

More explicitly, in \cite{nonstandard} the following configuration was considered: two SM2-branes located at  $(x_1, x_2, x_3)$ and $(x_4, x_5, x_6)$ and an SM5-brane located at $(x_1,..., x_6)$.
The metric and the 4-form field strength are
\bea
\la{metric}
ds^2 &=&-e^{2A}dt^2\,+\,
e^{2C_1}\,(dx_1^2+dx_2^2+dx_3^2)+\,e^{2C_2}\,(dx_4^2+dx_5^2+dx_6^2)
+\,e^{2D}\,d\Sigma_{4,\sigma}^2 \, ,  \\
F &=& P(t) e^{6C_1}\, dt\wedge dx_1\wedge dx_2\wedge dx_3 + 
R(t)e^{6C_2} \, dt\wedge dx_4\wedge dx_5\wedge dx_6 +
q \, \textrm{Vol}(\Sigma_{4,\sigma}) 
\label{4form}
\eea
where $d\Sigma_{4,\sigma}^2$ is the metric of the $4$-dimensional unit sphere ($\sigma=1$), unit hyperbola ($\sigma=-1$) or flat space ($\sigma=0$) and $\textrm{Vol}(\Sigma_{4,\sigma})$ is its volume form. The constant $q$ is proportional to the charge of the SM5-brane and metric functions depend only on time $t$. Note that (\ref{4form}) satisfies the Bianchi identity $dF=0$ trivially. We use our freedom of choosing the time coordinate to fix 
\be
A= 3C_1+3C_2+4D \, ,
\la{gauge1}
\ee
which simplifies the Ricci tensor and  introduce the function $G(t)$ through the relation
\be
D=G-C_1-C_2\, .
\la{gauge2}
\ee
Therefore, there are 5 unknown functions of time which are  $P$, $R$, $C_1$, $C_2$ and $G$. 

After these, the field equations take the form \cite{nonstandard}:
\bea
\la{form1}
P'&=& q\, R\, e^{6C_2} \, ,\\
\la{form2}
R'&=& - q\, P\, e^{6C_1} \, ,\\
\la{eqn1}
C_1'' &=& -\frac{1}{3}P^2\, e^{6 C_1} + \frac{1}{6} R^2 \, e^{6C_2} - 
\frac{q^2}{6} \, e^{6C_1+6C_2} \, ,\\
C_2'' &=& \frac{1}{6}P^2\, e^{6 C_1} - \frac{1}{3} R^2 \, e^{6C_2} - 
\la{eqn2}
\frac{q^2}{6} \, e^{6C_1+6C_2} \, , \\
G''&=& -3\, \sigma \, e^{6G} \, ,\label{g0}
\eea
and 
\be
2A'^2-6C_1'^2-6C_2'^2-8D'^2=P^2e^{6C_1} + R^2 e^{6C_2}
+ q^2e^{6C_1+6C_2},\label{cons}
\ee
where all derivatives are with respect to the time coordinate $t$. In the above system,  the first two equations come from the 4-form field equation (\ref{eins2}) and the following three  (\ref{eqn1})-(\ref{g0}) arise from the spatial components of the Ricci tensor (\ref{eins1}) which is diagonal. The last equation comes from the time component of the Ricci tensor (\ref{eins1}), which can be viewed as a constraint for initial data. When $q=0$, these equations correspond to those of a {\sl non-standard} 
SM2 $\perp$ SM2(-1) intersection 
whose solution was obtained in \cite{nonstandard} with $P^2=R^2=constant \neq 0$. Here {\sl non-standard} refers to S-brane intersections without supersymmetric $p$-brane analogs and (-1) means that SM2's have no common directions. When $P=R=0$ they reduce to those of an SM5-brane \cite{s1} and when $P=constant \neq 0$ and $R=q=0$ they correspond to an SM2-brane \cite{s2,s3}. Notice that equations remain unchanged if we interchange $ P \leftrightarrow R, C_1 \leftrightarrow C_2$ and  $q \leftrightarrow -q$.

To solve this system we first note that  (\ref{g0}) can be integrated once to give,
\be
(G')^2+\sigma e^{6G}= m^2 , 
\label{g2}
\ee
where $m$ is a constant. We obtain its solutions as 
\be
\label{G}
e^{-6G}=
\left\{
\begin{array}{ccc}
m^{-2}\sinh^2\left[3m\,(t-t_0)\right],
 \,\,\, &\sigma=-1& \,\,\, \textrm{(hyperbola)}, \\
m^{-2}\cosh^2\left[3m\,(t-t_0)\right],
 \,\,\, &\sigma=1& \,\,\, \textrm{(sphere)}, \\
\exp[6m\,(t-t_0)], \,\,\,  &\sigma=0& 
\,\,\, \textrm{(flat)},
\end{array}
\right. 
\ee
where $t_0$ is a constant. Using (\ref{G}) in (\ref{cons}) the constraint equation becomes
\be
24 m^2 = P^2e^{6C_1} + R^2 e^{6C_2}
+ q^2e^{6C_1+6C_2} + 12(C_1')^2 + 12(C_2')^2 + 12C_1'C_2' \, .
\la{integ}
\ee
Therefore, $G$ completely decouples from the system and we now have 4 equations (\ref{form1})-(\ref{eqn2}) and a constraint (\ref{integ}) for the functions $P$, $R$, $C_1$ and $C_2$. 

To proceed, it is useful to notice 
\be
2q(C_1'' - C_2'')= (PR)'\, , 
\ee
which implies
\be
2q(C_1'-C_2')=PR+ \tilde{e} \, , 
\la{simple}
\ee
where $\tilde{e}$ is an integration constant. We will use this simpler equation instead of (\ref{eqn2}) without loss of generality. Now, our strategy is to express $C_1$ and $C_2$ from (\ref{form1})-(\ref{form2}) in terms of $P, P', R$ and $R'$ and then use these in (\ref{eqn1}) and (\ref{simple}). As a result, we get the following two differential equations for $R$ and $P$:
\bea
\la{eqn3}
\frac{R'}{R} \,  \left(\frac{P'}{P} + \frac{PR}{q} \right) &=& \left(\frac{R''}{R'} - \frac{P'}{P} - \frac{PR}{q}\right)' \, , \\
\la{eqn4}
\frac{3PR}{q} +e &=& \frac{R'}{R} -\frac{P'}{P} + \frac{R''}{R'} - \frac{P''}{P'} \,\,\,  ,
\eea
where $e= \tilde{e}/q$. The integrability condition (\ref{integ}) becomes
\be
24m^2= \frac{RP'}{q} - \frac{PR'}{q} -\frac{P'R'}{PR} + \frac{1}{3}(\frac{R''}{R'}-\frac{P'}{P})^2 + \frac{1}{3}(\frac{P''}{P'}-\frac{R'}{R})^2 + \frac{1}{3}(\frac{R''}{R'}-\frac{P'}{P})(\frac{P''}{P'}-\frac{R'}{R}).
\la{integ2}
\ee
To summarize, at this stage our problem is reduced to solving two coupled differential equations 
(\ref{eqn3})-(\ref{eqn4}) subject to the condition (\ref{integ2}). After these 
one can read $e^{6C_1}$ and $e^{6C_2}$ from (\ref{form1})-(\ref{form2}) and determine $A$ and $D$ using (\ref{gauge1}) and (\ref{gauge2}).

To solve this complicated system we try the following ansatz in (\ref{eqn3})
\be
\frac{P'}{P} + \frac{PR}{q} = cR \, , 
\la{ansatz}
\ee
where $c$ is a constant. With this substitution we derive from (\ref{eqn3})
\be
\frac{R''}{R'} = 2cR+2d = 2\left( \frac{P'}{P} + \frac{PR}{q} \right) + 2d\, , 
\la{r''}
\ee
which gives a Riccati type differential equation for $R$:
\be
R'-cR^2-2dR=b \, , 
\la{diff}
\ee
where $d$ and $b$ are integration constants. The solution of this equation depends on the  combination $d^2-bc$. Postponing the explicit form of $R$ for the moment, we observe that the function $P$ can be solved either from (\ref{eqn4}) or (\ref{ansatz}). Substituting (\ref{r''}) in  (\ref{eqn4}) one gets
\be
 P^2R e^{(3d-e)t}=aP'\sqrt{R'}  \, ,
\la{p'}
\ee
where $a$ is another constant. Using (\ref{p'}) in our ansatz (\ref{ansatz}), $P$ can algebraically be solved as
\be
P=\frac{aqc\sqrt{R'}}{a\sqrt{R'}+q e^{(3d-e)t}} \, . \label{p}
\ee
Now, it only remains to check the integrability condition (\ref{integ2}) which gives
\be
24m^2= \frac{4d^2}{3} - bc \,\,  \hspace{1cm}  \textrm{and} \hspace{1cm} e=2d.
\la{constants}
\ee
From (\ref{form1})-(\ref{form2}) using (\ref{diff}), (\ref{p}) and (\ref{constants}) we find
\bea
\la{c1}
e^{6C_1} &=& -\frac{\sqrt{R'}\, (a\sqrt{R'}+qe^{dt})}{acq^2} \, , \\
\la{c2}
e^{6C_2} &=& \frac{aqc^2\sqrt{R'} \, e^{dt}}{(a\sqrt{R'} + q e^{dt})^2} \, .
\eea
Of course, the right-hand sides of these two equations should be non-negative for all $t$
which require:
\be
i) \, c<0 \, , \hspace{1cm} ii)\,  R'\geq 0 \, , \hspace{1cm} iii)\, aq>0 \, .
\la{conditions}
\ee 
Looking at the differential equation (\ref{diff}) we see that the first two of these requirements are satisfied only when 
$d^2>bc$. In this case there are two different solutions of (\ref{diff}) which read
\be
R=\begin{cases}{ -\frac{k^2}{c} \tanh [k^2 \, (t-t_1)] - \frac{d}{c}, \cr\cr -\frac{k^2}{c} \coth [k^2 \, (t-t_1)] - \frac{d}{c},}\end{cases}
\la{r21}
\ee
where $k^2 \equiv \sqrt{d^2-bc}$ \, and \, $t_1$ is a constant. However, conditions in (\ref{conditions}) choose the tangent hyperbolic function in (\ref{r21}) and we get
\be
R=-\frac{k^2}{c} \tanh [k^2 \, (t-t_1)] - \frac{d}{c},
\ee
which completes the solution process.

Before we study properties of this solution let us point out that one can try to generalize the ansatz (\ref{ansatz}) by replacing $cR$ with $cR^n$, where $n$ is a constant. Again solutions for $R$ and $P$ are easily obtained. However, one finds that the integrability condition (\ref{integ2}) is satisfied only for $n=1$. Let us also indicate that the equation (\ref{eqn3}) can be written as
$$
\frac{P'}{P} \, \left(\frac{R'}{R}-\frac{PR}{q}\right)=\left( \frac{R''}{R'} - \frac{P'}{P}-\frac{2PR}{q}\right) \, ,
$$
from which one can proceed by choosing 
$$
\frac{R'}{R}-\frac{PR}{q} = \tilde{c}P \, ,
$$
where $\tilde{c}$ is a constant. However, this leads to a solution which can be found  from the above by replacing $ P \leftrightarrow R, C_1 \leftrightarrow C_2$ and  $q \leftrightarrow -q$  which corresponds to the symmetry of the field equations that we mentioned earlier.

\subsection{Basic Properties}
\la{basic}

One may wonder if there are some redundant integration constants in the solution. By defining new coordinates $t\to t/c$ and $x_{4,5,6}\to c^{-1/3}x_{4,5,6}$ and further scaling $k^2\to -c k^2$, $m \to cm$ and $d\to cd$, one can remove the constant $c$ from the solution. Similarly, by scaling $x_{1,2,3}\to k^{-2/3}x_{1,2,3}$, $t\to t/k^2$, $m\to k^2 m$ and $d\to k^2 d$, it is possible to set $k=1$. Among other constants it is only allowed to set $\{d, t_0, t_1\}$ to zero in the solution. This means that SM2 $\perp$ SM2(-1), single SM2-brane and single SM5-brane solutions cannot be attained from ours by setting some constants to zero; it is intrinsically different and belongs to a different class. 
To analyze its characteristics further, let us we rewrite the solution after these scalings as 
(below we call the constant $a/q>0$ as $e^{t_2}$)
\bea
ds^2 &=&-e^{2A}dt^2\,+\,
e^{2C_1}\,(dx_1^2+dx_2^2+dx_3^2)+\,e^{2C_2}\,(dx_4^2+dx_5^2+dx_6^2)
+\,e^{2D}\,d\Sigma_{4,\sigma}^2\, , \nonumber\\
F &=& P(t) e^{6C_1}\, dt\wedge dx_1\wedge dx_2\wedge dx_3 + 
R(t)e^{6C_2} \, dt\wedge dx_4\wedge dx_5\wedge dx_6 +
q \, \textrm{Vol}(\Sigma_{4,\sigma}), \nonumber
\eea
where 
\be
A= 3C_1+3C_2+4D,\hspace{1cm} D=G-C_1-C_2\, ,
\ee
the function $G$ is given in (\ref{G}) and
\bea
R &=& -\tanh [t-t_1] - d\, , \nonumber\\
P &=& q\left(1+ e^{dt-t_2} \cosh [t-t_1]\right)^{-1}  , \nonumber\\
e^{6C_1} &=& \frac{1+ e^{dt-t_2} \cosh [t-t_1]}{q^2\cosh ^2 [t-t_1]} \, ,\nonumber\\
e^{6C_2} &=&  \frac{e^{dt-t_2}\cosh [t-t_1]}{(1+ e^{dt-t_2} \cosh [t-t_1])^2} \, ,
\la{solution}
\eea
with the restriction
\be\label{cons1}
24m^2=\frac{d^2}{3}+1\, .
\ee
The solution above is characterized by 6 constants  $\{m,d, q, t_0, t_1,t_2\}$. The condition (\ref{cons1}) relates $m$ and $d$, and it is possible to set one of the time constants $(t_0,t_1)$ to zero (in this section we set $t_0=0$ below) by shifting $t$  and redefining $t_2$. Therefore, there are actually 4 independent parameters which is less than the number of constants that appear in  {\sl standard} intersections of three SM-branes \cite{sinter1}. In those, there are 9 integration constants and only one of them can be set to 1.

It is easy to see that for any choice of the free parameters, there are curvature singularities as $t\to\pm\infty$. In these two limits, if $e^{dt}\cosh(t)\to \{0,  \infty \}$ then $e^{C_2}\to0$. On the other hand if $e^{dt}\cosh(t)\to1$, i.e. if $d=\pm1$, this time $e^{C_1}\to0$. Thus, either $(x_1,x_2,x_3)$ or $(x_4,x_5,x_6)$ spaces collapses as $t\to \pm\infty$, producing genuine curvature singularities. Nevertheless, the charge functions $P$ and $R$ are always finite and they approach to constants as $t \to \pm \infty$. Let us also note that when $d=0$ and $e^{-t_2}=|q|$ the function $C_1$ becomes equal to $C_2$ as $t \to \pm \infty$ and the metric of the solution approaches to the metric of an SM5-brane.

In the hyperbolic solution ($\sigma=-1$), the time coordinate $t$ is defined in the positive real line $t\in (0,+\infty)$. Defining the proper time 
\be
d\tau=e^A dt,
\ee
one can see that $\tau\sim t^{-1/3}$ as $t\to0$. Therefore, $t\to 0$ corresponds to the asymptotic region having an infinite proper time distance. In this limit, the metric approaches to the flat space
\be
ds^2\to-d\tau^2+\tau^2d\Sigma_{4,-1}^2+(dx_1^2+..+dx_6^2).
\ee 
On the other hand, for any choice of the parameters $m$ or $d$, the proper time converges to a finite value as $t\to\infty$. Hence, the hyperbolic solution represents a singular big-bang occurred at $t=\infty$ evolving to a flat space asymptotically as $t\to0$. 

In the spherical solution ($\sigma=1$),  $t$ is defined in the whole real line $t\in (-\infty,+\infty)$. One can deduce that as $t\to\pm\infty$, $e^A$ vanishes exponentially for any choice of the parameters, which implies that these two limits are actually separated by a finite proper time. This solution represents evolution from a big-bang to a big-crunch 

In the flat solution ($\sigma=0$),  $t$ is again defined in the whole real line $t\in (-\infty,+\infty)$. For $m>0$, one can see that $t\to\infty$ corresponds to a finite proper time interval but as $t\to-\infty$ the proper time diverges. Thus, for this case there is an initial big-bang singularity at $t=\infty$ but the big-crunch is an infinite proper time away from big-bang. If $m<0$, the roles played by the infinities change, i.e. while $t=-\infty$ corresponds to the big-bang, $t\to\infty$ labels the infinitely distant big-crunch. 

\subsection{Smearing and Dimensional Reduction to $D=10$}

In usual S-brane solutions it is possible to smear some directions along the $\S$-manifold \cite{sinter1}
until the overall transverse space is two dimensional. For the Chern-Simons S-brane solution  this again turns out  to be doable. We smear one direction by changing the transverse part of the metric (\ref{metric}) as follows:  
\be
\,e^{2D}\,d\Sigma_{4,\sigma}^2\to e^{2E}dy^2+\,e^{2{\hat D}}\,d\Sigma_{3,\sigma}^2 \, .
\ee
The 4-form field strength (\ref{4form}) corresponding to the SM5-brane is also modified,
\be
q \,  \textrm{Vol}(\Sigma_{4,\sigma}) \to
q \,  \textrm{Vol}(\Sigma_{3,\sigma}) \wedge dy \, .
\la{4form2}
\ee
One can check that the equations and hence the solutions (\ref{solution}) for $P$, $R$, $C_1$ and $C_2$ do not change under these modifications  if  $t$-reparametrization invariance is fixed by imposing
\be
{\hat A}=3C_1+3C_2+3{\hat D}+E.
\ee
The field equation for $E$ implies that
\be
E=\l t - t_3 -C_1-C_2,
\ee
where $\l$ and $t_3$ are constants. Introducing the function $\hat{G}$ as before
\be
{\hat D}={\hat G}-C_1-C_2,
\ee
it again decouples from the system and can be determined as 
\be
4{\hat G}= 6G-2\l t + 2t_3\, ,
\la{Gn}
\ee
where the function $G$ is given in (\ref{G}). Finally, the constraint equation becomes
\be
12m^2=3\l^2+\fr{d^2}{3}+1 \, .
\ee
This smearing is especially interesting, since it allows one to compactify along $(x_1,...,x_6,y)$ 
which gives flat, spherical or hyperbolic Robertson-Walker type cosmologies in $D=4$ as we will do in the next section. 

We can also use the smeared $y$-coordinate for direct dimensional reduction of our solution to type IIA theory by applying the formula
\be
ds^2_{11}= e^{-\phi/6}ds^2_{10, E} + e^{4\phi/3} dy^2 \, .
\la{reduction}
\ee
Then, the metric of the $D=10$ solution  in the Einstein frame is
\be
ds^2_{10, E} =  e^{\phi/6}[-e^{2{\hat A}}dt^2 + e^{2C_1}(dx_1^2+dx_2^2+dx_3^2) + e^{2C_2}(dx_1^4+dx_5^2+dx_6^2)
+ e^{2{\hat D}}d\Sigma_{3,\sigma}] \, ,
\la{metric2}
\ee
where the dilaton $\phi$ is given by
\be
\phi = \frac{3}{2}E = \frac{3}{2} \l t-\frac{3}{2}t_3 - \frac{1}{4} \ln \left[\frac{e^{dt-t_2}}{q^2 \cosh[t-t_1](1+e^{dt-t_2} \cosh[t-t_1])} \right] \, .
\ee
From the reduction of the 4-form field strength (\ref{4form}) with the modification (\ref{4form2}) we see that, we now have a solution (\ref{metric2}) with two SD2-branes located at $(x_1,x_2,x_3)$ and $(x_4,x_5,x_6)$ and an SNS5-brane located at $(x_1,...,x_6)$.

\section{Compactification to $D=4$ and Accelerating Cosmologies}

It is well-known that compactified SM2-branes can yield accelerating cosmologies in 4-dimensions \cite{acc1}-\cite{acc5}. Although the number of e-foldings is not large enough to use these solutions as realistic models of inflation,  they might be useful in understanding the current observed acceleration \cite{late}. In any case, keeping in mind that it is hard to obtain acceleration in string/M theory such solutions are worth to construct and study. Since in all previously obtained S-brane solutions in the literature $F\wedge F$ term trivially vanishes, one may ask whether a solution with $F\wedge F\not=0$ still gives accelerating cosmologies in 4-dimensions. Our aim in this section is to answer this question. 

Consider a metric in $(d+n)$-dimensions that has the form of a warped compactification
\be
ds_{d+n}^2=ds_d^2+\sum_i e^{2F_i} ds_i^2,
\ee
where $F_i$ are functions defined in $d$-dimensions. Then, the $d$-dimensional Einstein frame metric is given by
\be\label{compr}
ds_E^2=e^{\fr{2}{d-2}\sum_iF_i}ds_d^2 \, .
\ee
For the Chern-Simons S-brane solution, there are 3 different ways of compactifying to 4-dimensions: it is possible to reduce the solution along $(x_1,x_2,x_3,\S_4)$ or $(x_4,x_5,x_6,\S_4)$. Moreover, after smearing one direction in the transverse space $\S_4\to y \oplus \S_3$, we have an extra option of compactifying along $(x_1,...,x_6,y)$. One can see that in all these different possible compactifications, the 4-dimensional Einstein metric takes the form 
\be
ds_E^2=-S^6\,dt^2+S^2\,ds_3^2,
\ee
where $S$ is a function of time which can be determined using (\ref{compr}). Recalling that the proper time is given by $d\t=S^3dt$, the expansion and acceleration parameters can be found respectively as
\be
H=S^{-1}\fr{dS}{d\t}=S^{-4}\fr{dS}{dt},\hspace{1cm} {\tilde a}= \fr{d^2S}{d\t^2}=-\fr{1}{2}S^{-3}\fr{d^2}{dt^2}S^{-2}.
\ee
In an accelerating phase we demand $H>0$ and ${\tilde a}>0$. 

Let us first consider compactification along $(x_4,x_5,x_6,\S_4)$-directions. In that case the "scale factor" $S$ can be found as
\be
S=q^{1/3}\,e^{2G}\,\cosh^{1/4}(t)\,e^{-(dt-t_2)/12}.
\ee
For the flat solution a straightforward calculation shows that positive acceleration requires
\be
45+6(24m+d)\sinh[2(t-t_1)]-(24m+d)^2-[9+(24m+d)^2]\cosh[2(t-t_1)]>0.
\ee
In terms of $e^{2(t-t_1)}$, this inequality gives a quadratic equation which can easily be handled. We find that when the parameter $d$ is chosen in the interval
\be
0.2>d>-1.7,
\ee
there is a period of acceleration. During this period, the expansion speed can be made to be positive by a time reversal operation $t\to-t$. However, the number of e-foldings is of order unity, therefore the situation is not different than the usual S-brane solutions. 

For the hyperbolic and spherical cases, the expressions for the acceleration are much more complicated and it is difficult to perform an analytical examination. Using a numerical treatment, in the hyperbolic solution and for $t_0=t_1=t_2=0$, we observe that there is acceleration when $d$ is in the range
\be
0.4>d>-3.
\ee
In the spherical solution, acceleration turns out to be independent of the constant $t_2$. Setting $t_1=0$ by shifting time, we have two parameters $t_0$ and $d$ to adjust. We observe that for $t_0>0$ no acceleration occurs for any value of $d$. Choosing $t_0<0$,  positive acceleration can be obtained for a range of values of $d$ which varies with $t_0$. For instance, when $t_0=-1$ the acceleration takes place for
\be
0.2>d>-1.2.
\ee
The dependence of  the acceleration on a time shift parameter in the spherical case is a known phenomena that has been observed previously in \cite{acc4}. In all the cases described above, the number of e-foldings obtained during the accelerating phases are order unity.  

Consider next the compactification along $(x_1,x_2,x_3,\S_4)$-directions. The scale factor now becomes
\be
S=q^{1/6}\,e^{2G}\,[1+e^{dt-t_2}\cosh(t)]^{1/4}\,e^{-(dt-t_2)/6}.
\ee 
This time even for the flat solution the acceleration is very complicated and an analytical analysis is out of reach. By making plots for different sets of constants, we find that the 
accelerating phase can exist without the need of shifting the time constants
$(t_0,t_1,t_2)$ in the flat and hyperbolic solutions. For $t_0=t_1=t_2=0$, we observe acceleration in the flat background when
\be
2.4>d>0.3,
\ee
and in the hyperbolic solution when
\be
3>d>0.
\ee
In the spherical solution, the acceleration can only be obtained if time constants $(t_0,t_1,t_2)$ are chosen different. For instance, for $t_1=t_2=0$ and $t_0=5$ acceleration happens when
\be
-0.4>d>-2.5.
\ee

In such numerical plots, there is always the danger of missing the asymptotic behavior as $t\to\pm\infty$. In all the above  compactifications, the asymptotic structure of the 4-dimensional metrics can easily be determined in terms of the proper time. We find that while for the hyperbolic solution the metric asymptotically becomes   
\be
ds^2\to\begin{cases}{-d\t^2+\t^{2/3}d\S_3^2,\hspace{1cm}\t\to0,
\cr\cr-d\t^2+\t^{4/3}d\S_3^2,\hspace{1cm}\t\to\infty,}\end{cases}
\ee
in the spherical solution it can be written as 
\be
ds^2\to\begin{cases}{-d\t^2+(\t-\t_1)^{2/3}d\O_3^2,\hspace{1cm}\t\to\t_1,
\cr\cr-d\t^2+(\t-\t_2)^{2/3}d\O_3^2,\hspace{1cm}\t\to\t_2,}\end{cases}
\ee
where $\t_1$ and $\t_2$ are constants, and in the flat solution it goes like 
\be
ds^2\to\begin{cases}{-d\t^2+\t^{2/3}d\vec{x}^2,\hspace{1cm}\t\to0,
\cr\cr-d\t^2+\t^{2/3}d\vec{x}^2,\hspace{1cm}\t\to\infty.}\end{cases}
\ee
We therefore see that there is no acceleration in the asymptotic limits of these compactifications. 

Let us finally consider the compactification along $(x_1,...,x_6,y)$, where $y$ is a smeared direction in  $\S_4\to y \oplus \S_3$. In that case, the scale factor can be determined as
 \be
S=e^{\hat{G}+(\l t -t_3)/2},
\ee
where the function ${\hat G}$ is given in (\ref{Gn}). We thus have 
\be\label{sc3}
S=
\left\{
\begin{array}{ccc}
m^{1/2}\sinh^{-1/2}\left[2m\,(t-t_0)\right],
 \,\,\, &\sigma=-1& \,\,\, \textrm{(hyperbola)}, \\
m^{1/2}\cosh^{-1/2}\left[2m\,(t-t_0)\right],
 \,\,\, &\sigma=1& \,\,\, \textrm{(sphere)}, \\
\exp[-m\,(t-t_0)], \,\,\,  &\sigma=0& 
\,\,\, \textrm{(flat)}.
\end{array}
\right. 
\ee
Although using this compactification it is possible to obtain flat, hyperbolic and spherical Robertson-Walker type cosmologies in 4-dimensions, one can see from (\ref{sc3})  that accelerations are always negative for all three choices of $\sigma$.

\section{Corresponding Static Solutions}

In order to find static versions of our solution we need to change role played by the time coordinate with the radial coordinate in (\ref{metric}). After this, there are two alternatives for choosing the new time coordinate; it should either be in the initial transverse space $\Sigma_{4, \sigma}$ or should be chosen from the SM5-brane worldvolume. These are analogous to performing Wick rotations \cite{pope2}. The first option corresponds 
to a {\sl static} S-brane configuration whereas the second choice gives a non-extremal version of the composite M-brane configuration that was found in \cite{dyon1}.

\subsection{Version I: A Static S-brane Configuration}

It is known that there is a (nearly) one-to-one correspondence between S-branes and  {\sl static} timelike branes \cite{flux}. For the Chern-Simons S-brane it also turns out to be possible to find out the corresponding static solution. Assume the following metric and the 4-form field (note the minus sign in the last term in the form field in comparison to (\ref{4form})) 
\bea
ds^2 &=&e^{2A}dr^2\,+\,
e^{2C_1}(dx_1^2+dx_2^2+dx_3^2)+e^{2C_2}\,(dx_4^2+dx_5^2+dx_6^2)
+e^{2D}d\Sigma_{4,\sigma}^2 , \\
F &=& P(r) e^{6C_1}dr\wedge dx_1\wedge dx_2\wedge dx_3 + 
R(r)e^{6C_2}  dr\wedge dx_4\wedge dx_5\wedge dx_6 -
q \textrm{Vol}(\Sigma_{4,\sigma})
\eea
where all functions depend on $r$. Here $\Sigma_{4,\sigma}$ is the {\it Lorentzian} constant curvature four-manifold, that is the flat Minkowski, de Sitter and anti-de Sitter spaces for $\s=0$, $\s=+1$ and $\s=-1$ respectively whose Ricci tensors obey $R_{ij}=3\s g_{ij}$. Fixing the $r$-reparametrization invariance by 
\be
A=3C_1+3C_2+4D,
\label{param}
\ee
and introducing the  function $G$ as before
\be
D=G-C_1-C_2,
\label{g}
\ee
one can check that the unknown functions obey exactly the same set of differential equations, but this time as functions of $r$. Therefore, there is a solution given as
\be
e^{-6G}=
\left\{
\begin{array}{ccc}
m^{-2}\sinh^2\left[3m\,(r-r_0)\right],
 \,\,\, &\sigma=-1& \,\,\, \textrm{(anti-de Sitter)}, \\
m^{-2}\cosh^2\left[3m\,(r-r_0)\right],
 \,\,\, &\sigma=1& \,\,\, \textrm{(de Sitter)}, \\
\exp[6m\,(r-r_0)], \,\,\,  &\sigma=0& 
\,\,\, \textrm{(flat)},
\end{array}
\right. 
\ee
and 
\bea
R &=& -\tanh [r-r_1] - d\, , \nonumber\\
P &=& q\left(1+ e^{dr-r_2} \cosh [r-r_1]\right)^{-1}  , \nonumber\\
e^{6C_1} &=& \frac{1+ e^{dr-r_2} \cosh [r-r_1]}{q^2\cosh ^2 [r-r_1]} \, ,\nonumber\\
e^{6C_2} &=&  \frac{e^{dr-r_2}\cosh [r-r_1]}{(1+ e^{dr-r_2} \cosh [r-r_1])^2} \, ,
\eea
where constants obey
\be\label{cons2}
24m^2=\frac{d^2}{3}+1\, .
\ee
This is an example of a static solution in 11-dimensions for which $F\wedge F\not=0$. It can be interpreted as a system of three {\sl static} S-branes \cite{flux}: two static SM2 branes along $(x_1,x_2,x_3)$ and $(x_4,x_5,x_6)$ directions and a static SM5 brane along $(x_1,...,x_6)$.
Static S-branes do not preserve any supersymmetry \cite{flux}.

Unfortunately, the big-bang or the big-crunch singularities encountered in the time-dependent solutions become naked curvature singularities without a horizon. The flat and the spherical solutions become singular as $r\to\pm\infty$ and the hyperbolic solution is singular as $r\to\infty$.
As oppose to the time dependent background, now the hyperbolic solution
contains a naked singularity at $r=r_0$, since as $r \to r_0$ the metric approaches to
\be
ds^2\to d\tilde{r}^2+\tilde{r}^2(d\Sigma_{4,-1}^2)+(dx_1^2+...+dx_6^2),\hspace{1cm}\tilde{r}\to0.
\ee
Therefore, it is not possible to interpret these solutions as black objects. Since the asymptotic structure is not well defined due to the presence of singularities, it is also difficult to define proper conserved charges. 

\subsection{Version II: A Non-Extremal Composite M-brane Configuration}
\label{dyon}

There is another static version of our solution that resembles the composite M-brane solution of \cite{dyon1} which is a half supersymmetric, smooth solution. It was obtained using the U-duality transformations and it  consists of an $M2$-brane that lies inside an $M5$-brane and a {\sl static} SM2-brane. It reduces to $M2$ and $M5$ brane solutions for particular values of a constant. Intersections of this configuration was studied in \cite{dyon2}. Its anisotropic black  generalization \cite{black1} was obtained in \cite{dyon3} which has an additional function in front of the time coordinate in the metric. It is also possible to preserve the Poincar${\acute \textrm{e}}$ symmetry of a p-brane when finding its non-extremal version \cite{black2}. Our solution is related to such a generalization of the solution found in \cite{dyon1}. Let us assume the following metric and 4-form field
\bea
ds^2 &=&e^{2A}dr^2\,+\,
e^{2C_1}(-dt^2+dx_2^2+dx_3^2)+e^{2C_2}\,(dx_4^2+dx_5^2+dx_6^2)
+e^{2D}d\Sigma_{4,\sigma}^2 , \\
F &=& P(r) e^{6C_1}dr\wedge dt\wedge dx_2\wedge dx_3 + 
R(r)e^{6C_2}  dr\wedge dx_4\wedge dx_5\wedge dx_6 +
q \textrm{Vol}(\Sigma_{4,\sigma})
\eea
where all functions depend on $r$. Here $\Sigma_{4,\sigma}$ is the metric of the $4$-dimensional unit sphere ($\sigma=1$), unit hyperbola ($\sigma=-1$) or flat space ($\sigma=0$) and $\textrm{Vol}(\Sigma_{4,\sigma})$ is its volume form. So, we have an $M2$-brane located at $\{t, x_2,x_3\}$, an $M5$-brane at $\{t,x_2,...,x_6\}$ and a {\sl static} SM2-brane at $\{x_4,x_5,x_6\}$.
Fixing the $r$-reparametrization invariance  and the function $G$ as above  (\ref{param})-(\ref{g})
we get the following set of differential equations:
\bea
P'&=& q\, R\, e^{6C_2} \, ,\\
R'&=&  q\, P\, e^{6C_1} \, ,\\
C_1'' &=& \frac{1}{3}P^2\, e^{6 C_1} + \frac{1}{6} R^2 \, e^{6C_2} + 
\frac{q^2}{6} \, e^{6C_1+6C_2} \, ,\\
C_2'' &=& -\frac{1}{6}P^2\, e^{6 C_1} - \frac{1}{3} R^2 \, e^{6C_2} +
\frac{q^2}{6} \, e^{6C_1+6C_2} \, , \\
G''&=& -3\, \sigma \, e^{6G} \, ,
\eea
and 
\be
2A'^2-6C_1'^2-6C_2'^2-8D'^2 = -P^2e^{6C_1} + R^2 e^{6C_2}
- q^2e^{6C_1+6C_2},
\ee
where all derivatives are with respect to the radial coordinate $r$. Comparing with what we had for the time-dependent solution (\ref{form1})-(\ref{cons}) we see that they are quite similar. The new set can be obtained from the first by the transformation $P \to iP$ and $q \to iq$. Following the same strategy, 
we see that equations (\ref{g2})-(\ref{constants}) remain the same except the obvious replacement of $t$ with $r$. The equation (\ref{c2}) also remains the same and  the first difference appears in (\ref{c1}) where there is no more a minus sign on the right. This changes the condition on the integration constant $c$ in (\ref{conditions}) to $c>0$, others remaining the same. Because of this, the solution for $R$ is now given by the cotangent hyperbolic function in (\ref{r21}). In summary, our static solution is:
\bea
R &=& -\coth [r-r_1] - d\, , \nonumber\\
P &=& q\left(1+ e^{dr-r_2} \sinh [r-r_1]\right)^{-1}  , \nonumber\\
e^{6C_1} &=& \frac{1+ e^{dr-r_2} \sinh [r-r_1]}{q^2\sinh ^2 [r-r_1]} \, ,\nonumber\\
e^{6C_2} &=&  \frac{e^{dr-r_2}\sinh [r-r_1]}{(1+ e^{dr-r_2} \sinh [r-r_1])^2} \, ,
\eea
with the restriction
\be
24m^2=\frac{d^2}{3}+1\, .
\ee
The function $G$ is given as 
\be
e^{-6G}=
\left\{
\begin{array}{ccc}
m^{-2}\sinh^2\left[3m\,(r-r_0)\right],
 \,\,\, &\sigma=-1& \,\,\, \textrm{(hyperbola)}, \\
m^{-2}\cosh^2\left[3m\,(r-r_0)\right],
 \,\,\, &\sigma=1& \,\,\, \textrm{(sphere)}, \\
\exp[6m\,(r-r_0)], \,\,\,  &\sigma=0& 
\,\,\, \textrm{(flat)},
\end{array}
\right. 
\ee
through which we can read the remaining metric functions as $A = 4G - C_1 - C_2$ and $D = G - C_1- C_2$.
This is a non-extremal version of the solution obtained in \cite{dyon1} with a more general transverse space. It is not possible to remove $M2$ or $M5$-brane from the system in contrast to  \cite{dyon1}. Unlike \cite{dyon3} the $M2$-brane worldvolume is isotropic and there seems to be no obvious extremal limit.

\section{Conclusions}

Solutions of the $D=11$ supergravity \cite{11} with non-vanishing $F \wedge F$ are very rare. 
In this paper we have found three such examples. We hope that these novel solutions will be helpful in studying various effects of the Chern-Simons term. Our first example is an intersecting S-brane configuration. Being a time-dependent solution it is appropriate for cosmological applications. It does not reduce to known S-brane solutions by setting some constants to zero which shows that the Chern-Simons term plays an essential role. However, like usual S-branes
its metric is singular. Moreover, we also found two static versions of this. The first one does not have a $p$-brane interpretation but can be thought of as a {\sl static} S-brane configuration. The second static solution is a non-extremal version of the composite M-brane solution found in \cite{dyon1}. The connection of ours with \cite{dyon1} and its anisotropic non-extremal generalization \cite{dyon3} needs to be explored further.

A primary motivation of considering the above S-brane system was to see whether 
the Chern-Simons flux modifies acceleration obtained from SM2-branes after compactification to 4-dimensions. We found that there is again a period of acceleration in two of our compactifications. However, despite this new ingredient there is no dramatic change in 
expansion factors compared to usual SM-branes and it is still far from explaining the early universe inflation. 
It would be nice to understand this also from the effective theory obtained in 4-dimensions \cite{acc4}. 
For this the dimensional reduction of the Chern-Simons term is necessary which was obtained in \cite{reduction}. Actually this type of transient acceleration is quite a common 
feature of a large class of time-dependent M-theory compactifications as was shown in \cite{nogo1, 
nogo2}. It would be very interesting to investigate whether such  no-go theorems are still valid when the Chern-Simons term contributes. The relevance of our solution to the present day acceleration \cite{late} also remains to be seen.

There are several possible extensions of our work. For instance, one can consider curved worldvolumes for SM2-branes. Furthermore, one can try 
to add more S-branes or p-branes \cite{sp1, sp2} to the system. The first two generalizations might increase the acceleration rate and the last one  might be convenient for studying non-homogeneous cosmologies. Another interesting thing to do is to consider double dimensional reductions of our solutions to ten dimensions as was done for SM-branes \cite{roy}.
In addition to these, it is desirable to see whether our solution is unique. We plan to come back to these issues soon.

\begin{acknowledgments}
We are grateful to Miguel S. Costa for bringing the composite M-brane solution to our attention, which led us to find the second static solution presented in the subsection \ref{dyon}. NSD is partially supported by Turkish Academy of Sciences via The Young Scientists Award Program (T\"UBA- GEB\.IP). He also wishes to thank the Abdus Salam ICTP for hospitality where some part of this paper was written. 
\end{acknowledgments}

\end{document}